\documentclass[showpacs,preprintnumbers,amsmath,nofootinbib,aps]{revtex4}

\usepackage{graphicx,epsf, epsfig, psfig, amssymb}
\usepackage{bm}

\begin{document}

\title{Simulations of black hole air showers in cosmic ray detectors}
\author{Eun-Joo Ahn}
\email{sein@oddjob.uchicago.edu}
\affiliation{Department of Astronomy and Astrophysics,
and
Kavli Institute for Cosmological Physics,
The University of Chicago, 5640 S.\ Ellis Ave, Chicago IL 60637, USA}

\author{Marco Cavagli\`a}
\email{cavaglia@olemiss.edu}
\affiliation{Department of Physics and Astronomy, University of Mississippi,
University, MS 38677-1848, USA}

\date{\today}

\begin{abstract}

We present a comprehensive study of TeV black hole events in Earth's atmosphere
originated by cosmic rays of very high energy. An advanced fortran Monte Carlo
code is developed and used to simulate black hole extensive air showers from
ultrahigh-energy neutrino-nucleon interactions. We investigate the
characteristics of these events, compare the black hole air showers to standard
model air showers, and test different theoretical and phenomenological models
of black hole formation and evolution. The main features of black hole air
showers are found to be independent of the model considered. No significant
differences between models are likely to be observed at fluorescence
telescopes and/or ground arrays. We also discuss the tau ``double bang''
signature in black hole air showers. We find that the energy deposited in the
second bang is too small to produce a detectable peak. Our results show that
the theory of TeV-scale black holes in ultrahigh-energy cosmic rays leads to
robust predictions, but the fine prints of new physics are hardly to be
investigated through atmospheric black hole events in the near future.

\end{abstract}


\maketitle

\section{Introduction}
The study of super-Planckian collisions dates back to the late 80's
\cite{Amati:1987wq}. Today's renewed interest \cite{Banks:1999gd} stems from
the possibility that the fundamental scale of gravity may be much lower than
the observed gravitational scale \cite{Arkani-Hamed:1998rs}. In braneworld
scenarios, the observed weakness of the gravitational field is due to the
``leakage'' of gravity in the extra dimensions: Standard model (SM) fields are
constrained in a four-dimensional submanifold, whereas gravitons are allowed to
freely propagate in the higher-dimensional spacetime \cite{Maartens:2003tw}. If
the gravitational coupling constant is of the order of few TeVs, the physics of
super-Planckian collisions could soon be detected through observation of
subnuclear black holes (BHs) and other extended objects, such as branes, in
particle colliders \cite{Giddings:2001bu, Ahn:2002mj} or ultrahigh-energy
cosmic ray (UHECR) observatories \cite{Feng:2001ib, Anchordoqui:2001cg,
Ahn:2003qn,  Cardoso:2004zi}. (For reviews and extra references, see Refs.\
\cite{Cavaglia:2002si, Landsberg:2002sa, Cardoso:2005jq}).

The semiclassical limit of super-Planckian scattering suggests that the cross
section for creation of a BH or brane with radius $R$ is approximately given by
the geometrical black disk $\sigma_{BD}(s,n)=\pi R^2(s,n)$, where $\sqrt{s}$ is
the center of mass (c.m.) energy of the colliding quanta and $n$ is the number
of extra dimensions. Gravitational objects with mass of order of the
fundamental gravitational scale $M_\star$ have radius of order $M_\star^{-1}$.
In symmetric compactification models, the size of extra dimensions is much
larger than $M_\star^{-1}$. (For conventions, see Ref.\
\cite{Cavaglia:2002si}.)  Thus the spherical approximation is justified; the
geometry of nonperturbative objects is that of a $n$-dimensional BH. The
spherical approximation breaks down for asymmetric compactifications, where
some of the extra dimensions have size of order of the fundamental Planck
scale. In that case, the geometry of nonperturbative objects is that of strings
and branes \cite{Ahn:2002mj}. 

UHECRs are attractive because of their high c.m.\ energy. The nucleon-nucleon
cross section for formation of BHs and branes is very small compared to other
SM hadronic processes. The neutrino-nucleon cross section for BH or brane
formation may be higher than the cross section of the SM process, thereby
giving interest to neutrino interaction. Under the most favorable
circumstances, the cross section for BH formation at the TeV scale reaches
millions of pb for neutrino-nucleon collisions in the atmosphere. The cross
section for brane production is expected to be even larger. These results have
led to the claim that UHECR detectors might observe BHs and probe Planckian
physics. Event observables would be the secondary products of the BH or brane
after their formation, i.e.\ extensive air showers originated by field emission
in the decay phase. Hawking evaporation provides an emission mechanism for BHs
\cite{Hawking:1974sw}. SM quanta are emitted in the visible three-brane and can
be detected. Branes may or may not evaporate, depending on their properties.
However, the decay spectrum of massive excitations in string theories has been
shown to be thermal \cite{Amati:1999fv}. This suggests that BH and brane decay
signatures may be similar. 

The observational signatures of BH events in the semiclassical approximation
have been investigated in a number of recent publications. In
Ref.~\cite{Ahn:2003qn} the authors found that BH interactions generate
different air showers from SM interactions. BH air showers tend to rise faster
and have larger muon content. A BH event produces a hadronic air shower
occurring at a much greater depth in the atmosphere, i.e., a very deeply
penetrating hadronic air shower. However, the inability of realistic detectors
to observe the first interaction point hides most of the difference between BH
and SM air showers. Given that present observatories are not large enough to
study a large number of neutrino events, discrimination of BH and SM events is
likely not to be achieved in the near future. Another characteristic of BH air
showers is $\tau$ generation. Although the rate for these events is low if the
ultrahigh-energy neutrino flux is at the level of the expected cosmogenic
neutrinos, unusual ``double-bang'' air showers could signal a departure from SM
interactions.   

Whereas the semiclassical picture seems reasonable, the actual physics of
subnuclear BH formation could be very different. In the last year or two,
significant advances in the understanding of microscopic BH formation and air
shower evolution have appeared in the literature. It is thus timely and
worthwile to re-examine the observational signatures of BH air showers. To this
purpose, we developed a thorough fortran Monte Carlo (MC) code to simulate the
air showers induced by BH formation in neutrino-air collisions, which includes
these theoretical refinements \cite{Groke1.01:2005}. The MC has the same
structure of the MC used in Ref.\ \cite{Ahn:2003qn}. The code generates
observable secondaries from BH evaporation using the PYTHIA generator
\cite{Sjostrand:2000wi}. These secondaries are then injected into the AIRES
simulator \cite{aires} as primaries for the final air shower.   

The purpose of our study is threefold. Firstly, we want to confirm the main
findings of the previous investigation. Secondly, we want to test various
proposals of BH vs.~SM air shower discrimination that have appeared in the
literature, such as the $\tau$ ``double bang'' effect \cite{Cardoso:2004zi}.
Thirdly, we want to look for new ways of discriminating between different
models of BH formation and evolution.  

Our analysis will show that the main characteristics of BH air showers are
essentially independent of the details of BH evolution. Because of large
uncertainties and statistical fluctuations in air shower detection, it is also
practically impossible to discriminate between alternative models of BH
formation and evaporation. For instance, we will show that there is no
significant observational difference between a model of BH formation based on
the semiclassical black disk and the trapped-surface model
\cite{Cardoso:2005jq}, or between a model of BH evaporation with final
explosive decay and stable remnant \cite{Cavaglia:2004jw}. These results limit
significantly the use of BH air showers (if they exist) to probe details of
``new physics''. We will also show that newly proposed signatures do not help
in the task of discriminating BH vs. SM air shower detection. No observational
trace of the ``double bang'' signature can be extracted from a realistic
detector in the near future.
\section{Basics of BH formation and evolution}
In this section we briefly review the basics of BH formation and evolution,
focusing on recent theoretical advances that have been included in the MC
code.
\subsection{BH formation and cross section at parton level}
Thorne's hoop conjecture \cite{hoop} states that a horizon forms when a mass
$M$ is compacted into a region with circumference smaller than twice the
Schwarzschild radius $R(M)$ in any direction. At subnuclear level, this can be
achieved by scattering two partons $(ij)$ on the brane with  c.m.\ energy
$\sqrt{s_{ij}}>M$ and impact parameter $b<R(M)$. This event can formally be
described by the process $ij\to {\rm BH}+E(X)$, where $E(X)$ denotes
collisional energy that does not contribute to the BH mass. This energy
includes a bulk component of gravitational radiation and perhaps non-SM gauge
fields, and a brane component of SM fields. If $E(X)$ is zero, the hoop
conjecture implies that the cross section for BH production is independent of
the impact parameter (as long as $b<R(M)$) and equal to the geometrical black
disk $\sigma_{BD}(s_{ij},n)$. If $E(X)\not =0$, the cross section depends on
the impact parameter, and is expected to be smaller than the black disk cross
section. It is worth stressing that this picture is correct only if the BH is
larger than the Compton length of the colliding quanta. (For discussions on the
effect of wave packet size on the BH formation process, see Ref.\
\cite{Voloshin:2001vs}.) A precise calculation of the collisional energy loss
is essential to understand BH formation.  

Many papers have been devoted to improve or disprove the hoop conjecture. The
most popular model is currently the trapped-surface model
\cite{Yoshino:2002tx,Yoshino:2005hi,Vasilenko:2003ak}, although alternative
techniques have been explored \cite{Cardoso:2005jq}.  The trapped-surface
approach gives an upper bound on the gravitational component of $E(X)$ by
modelling the incoming partons as two Aichelburg-Sexl shock waves
\cite{Aichelburg:1970dh}. The Aichelburg-Sexl wave is obtained by boosting the
Schwarzschild solution to the speed of light at fixed energy. The resulting
metric describes a plane-fronted gravitational shock wave corresponding to the
Lorentz-contracted longitudinal gravitational field. The parton scattering is
simulated by superposing two shock waves travelling in opposite directions. The
union of these shock waves defines a closed trapped-surface that allows to set
a lower bound on the BH mass. The collisional energy loss depends on the impact
parameter and increases as the number of spacetime dimensions increases. The BH
mass monotonically decreases with the impact parameter from a maximum of about
60-70\% of the c.m.\ energy for head-on collisions.

The trapped-surface result is consistent within one order of magnitude with the
hoop conjecture. However, the partons are assumed to be pointlike, massless,
spinless, and electrically neutral. The pointlike assumption fails for
directions transversal to the motion \cite{Kohlprath:2002yh}. Colliding partons
generally have spin and charge. While size and spin effects are expected to be
mostly relevant around the Planck energy, charge effects could dominate at
higher energy. It should also be kept in mind that the trapped-surface model
provides only a lower bound on the BH mass. An accurate estimate of the
gravitational collisional energy loss would require the use of the full
non-linear Einstein equations in higher dimensions. Since this is a virtually
impossible task, alternative approximated models have been investigated. 
The gravitational energy emission in a hard instantaneous collision can be
computed in the linearized limit \cite{Cardoso:2002pa}. This approach suggests
that the trapped surface method overestimates the gravitational energy emitted
in the process. For head-on collisions, the instantaneous method predicts the
gravitational energy loss to be only about 10\% of the c.m.\ energy. This
result is in agreement with a perturbative calculation modelling the
parton-parton collision as a plunge of a relativistic test particle into a BH
with mass equal to the c.m.\ energy \cite{Berti:2003si}.  

In conclusion, known methods for the estimate of the gravitational loss in
relativistic scattering at parton level give a BH mass ranging between 60\% and
100\% of the c.m.\ energy. Today, the trapped-surface value and the black disk
value can be considered as the lower and upper bounds on the BH mass,
respectively.
\subsection{Cross section at nucleon level}
The total cross section for a super-Planckian event involving a nucleon is
obtained by integrating the above cross section over the parton distribution
functions. BHs formed in a neutrino-nucleon collision may dominate over the SM
processes and stand a fair chance of detection. On the contrary, the branching
ratio of the BH cross section in a nucleon-nucleon collision is $\sim
10^{-9}$. Therefore, BH detection in nucleon-nucleon interactions cannot be
achieved with current and next generation detectors due to the low flux of
UHECRs.

If the BH mass depends on the impact parameter, the generally accepted formula
for the total cross section of the neutrino-nucleon process is
\begin{equation}
\sigma_{\nu N \to BH} = \sum_{i}\int_0^1 2z dz\int_{x_m}^{1} dx
\, q_i(x,-Q^2) \, F\,\sigma_{BD}(xs,n)\,,
\label{totcross}
\end{equation}
where $q_i(x,-Q^2)$ are the Parton Distribution Functions (PDFs)
\cite{Brock:1993sz,Eidelman:2004wy} with four-momentum transfer squared $-Q^2$,
and fraction of the nucleon's momentum carried by the $i$-th parton $\sqrt{x}$.
$z$ is the impact parameter normalized to its maximum value and
$x_m=M_{min}^2/(sy^2(z))$, where $y(z)$ and $M_{min}$ are the fraction of
c.m.~energy trapped into the BH and the minimum-allowed mass of the
gravitational object, respectively. $F$ is a form factor. The total cross
section for the black disk model is obtained by setting $F=1$ and $y^2(z)=1$. 

Different sets of PDFs are defined in the literature. The PDFs are not known at
energies above the TeV and for values of momentum transfer expected in BH
formation. Equation (\ref{totcross}) is usually calculated by imposing a
cut-off at these values. The PDFs also suffer from uncertainties at any
momentum transfer ($\sim$ 10\%) \cite{Ahn:2003qn} and from the ambiguity in the
definition of $Q$ \cite{Emparan:2001kf}. The momentum transfer is usually set
to the BH mass or the inverse of the Schwarzschild radius. Although recent
literature inclines toward the latter, there are no definite arguments to
prefer either one or to exclude alternative choices. The uncertainty due to the
ambiguity in the definition of the momentum transfer is about $\sim 10 - 20$\%
\cite{Anchordoqui:2001cg}.

The form factor and the amount of trapped energy depend in principle on energy,
gravitational scale, geometry and physical properties of the extra dimensions
and gravitational object. The trapped-surface method gives numerical values of
order unity for these quantities. (See Refs.\ \cite{Yoshino:2002tx,
Yoshino:2005hi} and discussion above). However, these results depend on the way
the trapped-surface is identified. Other models \cite{Yoshino:2005ps} give
values which are more or less consistent with the trapped-surface method. With
the lack of further insight, it is common practise in the literature to either
choose the trapped-surface result or the simple black disk model. 

The lower cutoff on the fraction of the nucleon momentum carried by the partons
is set by the minimum-allowed (formation) mass of the gravitational object,
$M_{min}$. This threshold is expected to roughly coincide with the mass for
which the semiclassical description is valid. This conclusion is motivated for
spherically symmetric BHs by the following argument \cite{Anchordoqui:2003jr}:
For $M_{min}/M_\star\gtrsim$ few, the Hawking entropy of the BH should be large
enough to neglect strong gravitational effects. The semiclassical results are
then extrapolated for smaller values with the assumption that the BH or its
Planckian progenitor decays on the brane. However, this argument is based on
Hawking's semiclassical theory and may not be valid at energies equal to few
times the Planck mass. For example, the existence of a minimum spacetime length
$l_{m}$ implies the lower bound on the BH mass \cite{Cavaglia:2003qk,
Cavaglia:2004jw}:
\begin{equation}
M_{ml}= \frac{n+2}{8\Gamma\left(\frac{n+3}{2}\right)}
\left(2\sqrt{\pi}\,l_{m}/M_\star\right)^{n+1} \, M_{\star}\,.
\label{minmass}
\end{equation}
BHs with mass less than $M_{ml}$ do not exist, since their horizon radius would
fall below the minimum-allowed length. At fixed $M_\star$, the minimum-allowed
mass grows as a power of $l_m^{n+1}$. For $n=6$ or $7$ and $l_{m}M_\star\gtrsim
1$, it follows $M_{ml}\sim M_{min}\gg M_\star$.
\subsection{BH evolution}
It is believed that the decay of microscopic BHs happens in four distinct
stages: I. radiation of excess multipole moments (balding phase); II.
spin-down; III. Hawking evaporation; IV. final explosion or formation of a BH
remnant. Phases I-III rely on semiclassical results, provided that the entropy
is sufficiently large. Phase IV is in the realm of quantum gravity.  

Although some progress has been made, the understanding of balding phase and
spin-down phase is still fragmentary. For example, the emission of radiation
from a $(n+4)$-dimensional rotating BH on the brane is known only for spin-0
fields \cite{Duffy:2005ns}. Due to these limitations, some of these theoretical
results cannot be implemented in MC simulations at the present stage. Moreover,
phase I is not expected to lead to a significant amount of energy loss and the
observational uncertainties (see below) are likely to dominate the theoretical
uncertainties in phase II.

Many papers have been devoted to the investigation of the Hawking phase.
Although several analytical and numerical results have been obtained
\cite{Kanti:2002nr}, from the viewpoint of numerical simulations the
situation is similar to the balding and spin-down phases. For instance,
greybody factors for the graviton are not fully known in $(n+4)$-dimensions
even for the spherically symmetric BH. This precludes their use in numerical
codes, where a consistent use of greybody factors is required. The field
content at trans-Planckian energies is also not known. Onset of supersymmetry,
for example, could lead to other evaporation channels for the BH and large
emission of non-SM or undetectable quanta during the decay phase. Finally,
quantum effects may also affect the emission of visible quanta on the brane. 

Quantum corrections to the Hawking phase can be phenomenologically described by
assuming the existence of a minimum length of the order of the Planck length
\cite{Garay:1994en}. The existence of a minimum scale is a common consequence
of most (if not all) theories of quantum gravity such as string theory,
non-commutative geometry and loop quantum gravity. The presence of a cutoff at
small spacetime distances leads to a modification of the uncertainty principle
at Planck scales. Since the Hawking thermodynamical quantities can be derived
by applying the uncertainty principle to the BH, the existence of a minimum
length leads to corrections in the thermodynamical quantities
\cite{Cavaglia:2003qk,Cavaglia:2004jw}. 

At the end of the Hawking phase, the BH is expected to either non-thermally
decay in a number $n_p$ of hard quanta or leave a remnant. In either case we
must content ourselves with a phenomenological description, due to the lack of
a theory of quantum gravity. The final non-thermal decay is usually described
by setting a cutoff on the BH mass of the order of the Planck mass,
$Q_{min}\sim M_\star$, and then equally distributing the energy $Q_{min}$ to a
number $n_p$ of quanta. Since the decay is non-thermal, and in absence of any
guidance from a theory of quantum gravity, the quanta are democratically chosen
among the SM degrees of freedom. Note that $Q_{min}$ does not necessarily
coincide with $M_{min}$. The former gives the threshold for the onset of
quantum gravity effects, whereas the latter gives the minimum-allowed mass of
the classical object. From the above definitions, it follows $M_{min}\geq
Q_{min}$. The existence of a minimum length gives a natural means to set
$Q_{min}$. In that case, the modified thermodynamical quantities determine the
endpoint of Hawking evaporation when the mass of the BH reaches $M_{ml}$. This
mass can be identified with the mass of the BH remnant
\cite{Cavaglia:2003qk,Cavaglia:2004jw}.
\section{BH generator \label{BHgenerator}}
In this section we list the main characteristics of the MC generator used in
the simulations \cite{Groke1.01:2005}. The physics of BH formation and
decay is determined by the following set of external parameters and switches in
the MC code:
\begin{enumerate}
\item Fundamental Planck scale;
\item Number of extra dimensions;
\item Gravitational loss at BH formation and gravitational loss model;
\item Minimum BH mass at formation;
\item Quantum BH mass threshold at evaporation;
\item Number of final quanta at the end of BH decay;
\item Momentum transfer model in parton collision;
\item Conservation of electromagnetic (EM) charge;
\item Minimum spacetime length.
\end{enumerate}
The above parameters are briefly explained below. A more detailed explanation
can be found at the MC generator web site \cite{Groke1.01:2005}. 
\subsection{BH formation and parton cross section}
The MC does not require any lower or upper bound on the Planck mass $M_\star$.
However, experimental constraints exclude values of $M_\star\lesssim 1$ TeV
and large values of $M_\star$ do not allow BH formation in the atmosphere.
Therefore, $M_\star$ must be chosen with caution. Since $n=1$ and $n=2$ are
excluded experimentally, and most of the theoretical models are limited to
$n\leq 7$, the number of extra dimensions $n$ ranges from 3 to 7. 

The MC includes three models for BH formation and cross section: Black disk,
Yoshino-Nambu (YN) trapped-surface model \cite{Yoshino:2002tx}, and
Yoshino-Rychkov (YR) improved trapped-surface model \cite{Yoshino:2005hi}. This
allows a comparison between air showers based on the black disk model
\cite{Ahn:2003qn} and air showers generated by BHs with significant
gravitational loss at formation. Observable differences between different
models of BH formation can be investigated, as suggested in Ref.\
\cite{Yoshino:2005hi}.

The minimum BH mass $M_{min}$ is set in units of $M_\star$ or  $M_{ml}$ (if a
minimum length is present, see below). This parameter is always larger than
one, i.e.\ $M_{min}\geq{\rm Max}(M_\star,M_{ml})$. 
\subsection{Total and differential cross section}
The distribution of the initial BH masses is given by the differential cross
section $d\sigma/dM_{BH}$, where $M_{BH}=\sqrt{xs}$. The MC uses the (stable)
cteq5 PDF distribution \cite{Brock:1993sz,cteq5}. Since the use of different
PDF distributions produces an insignificant uncertainty in the total and
differential cross sections, other PDF distributions are not implemented in the
MC. The uncertainty due to the choice of the momentum transfer is generally
larger. Therefore, a switch allows to choose between BH mass or inverse of the
Schwarzschild radius as definition of momentum transfer. 

The part of c.m.\ energy of the neutrino-nucleon system which is not trapped or
lost in gravitational radiation at formation is attributed to the nucleon
remnant. For sake of simplicity, only neutrino-proton collisions are
implemented in the MC code. A neutrino-neutron collision does not produce
significant statistical differences in the nucleon remnant compared to a
neutrino-proton collision. The proton remnant is successfully fragmented
according to QCD in mesons/baryons (see, for example, Ref.\
\cite{Sjostrand:2000wi}) and then is decayed with the PYTHIA generator along
with the quanta created in the BH evaporation process.
\subsection{BH evaporation}
Due to the lack of results for the balding and spin-down phases described
above, energy losses in these stages are assumed to be either negligible or
included in the energy loss during formation. This is a reasonable assumption
since the trapped-surface model likely overestimates the actual energy loss. 
Balding and spin-down effects are also not expected to produce detectable
differences in BH air showers, given experimental uncertainties and statistical
fluctuations. Nevertheless, keeping an open mind, we plan to include balding
and spin-down effects in updated versions of the code, as soon as theoretical
results become available.

A similar conservative approach is used in the Hawking phase, where only
thermally-averaged greybody factors in four dimensions are implemented in the
MC. This is justified by consistency reasons in the code (the full greybody
factors for all fields are not known). As the SM fields are emitted on the
brane, and given the observational uncertainties, the difference between
thermally-averaged and exact greybody factors is not expected to be detectable.
The particle content at trans-Planckian energies is assumed to be the minimal
$SU(3)\times SU(2)\times U(1)$ SM with three families and a single Higgs boson.
The degrees of freedom $c_i$ and the thermally-averaged greybody factors
$\Gamma_i$ are listed in Table~I. The decay multiplicities per species $N_i$
are assigned according to the prescription of Ref.~\cite{Cavaglia:2003hg}:
\begin{equation}
N_i=N\,\frac{c_i\Gamma_{i}f_i(3)}{\sum_j
c_j\Gamma_{j}f_j(3)}\,,
\label{multii}
\end{equation}
where $f_i(m)=1$ or $1-2^{1-m}$ for bosons or fermions, and the total
multiplicity $N$ is
\begin{equation}
N=\frac{30\zeta(3)}{\pi^4}\,S\,\frac{\sum_i c_i\Gamma_{i}f_i(3)}
{\sum_j c_j\Gamma_{j}f_j(4)}\,,
\label{multin}
\end{equation}
where $S$ is the initial entropy of the BH. 

The presence of a minimum length affects the BH evolution in the Hawking phase.
If no minimum length is present, the MC evaporates the BH according to the
Hawking theory. Alternatively, the BH evolution proceeds according to the
modified thermodynamics of Ref.\ \cite{Cavaglia:2003qk,Cavaglia:2004jw}. In
both cases the evaporation ends when the BH reaches the mass $Q_{min}$. This is
set in units of $M_\star$ ($M_{ml}$) if the minimum length is zero (nonzero).
Note that the BH minimum formation mass $M_{min}$ and the endpoint of Hawking
evaporation $Q_{min}$ are independent parameters. 
\begin{table}[ht]
\begin{center}
\begin{tabular}{|l||c|c|}
\hline
& $c_i$ &  $\Gamma_i$ \\
\hline
 ~quarks & 72 &  ~0.6685~ \\
 ~charged leptons~ &  12 &  0.6685\\
 ~neutrinos &  6 &  0.6685 \\
 ~photon &  2 &  0.2404 \\
 ~EW bosons &   9 &  0.2404 \\
 ~gluons &  16 &  0.2404 \\
 ~Higgs &  1 &  1 \\
 ~graviton & ~$(n+4)(n+1)/2$~ &  0.0275 \\
\hline
\end{tabular}
\caption{Degrees of freedom $c_i$ and thermally-greybody factors $\Gamma_i$ for
the SM fields. The graviton is assumed to propagate in all dimensions.}
\end{center}
\label{emivalue}
\end{table}
Four-momentum is conserved at each step in the evaporation process by taking
into account the recoil of the BH on the brane due to the emission of the
Hawking quanta. The initial energy of the BH is distributed democratically
among all the Hawking quanta with a random smearing of $\pm 10$\%. This
smearing factor is introduced on a purely phenomenological basis to take into
account quantum uncertainties in the emission of each quantum.
\subsection{BH final decay}
The MC code allows for two different choices of final BH decay: Final explosion
in a number $n_p$ of quanta or BH remnant. If $n_p=0$, the BH settles down to a
remnant with mass $Q_{min}$. If $n_p$ = 1\dots 18, the BH decays in a number
$n_p$ of quanta by a $n$-body process with total c.m.\ energy equal to
$Q_{min}$. 

A switch controls conservation of EM charge in the decay process (Hawking
evaporation + final decay). The purpose of this switch is to allow for the
existence of a charged BH remnant. If the EM charge is not conserved and
$n_p=0$, the BH remnant carries a charge $Q_R$, where $1-Q_R=Q_H+Q_N$ is the sum
of the EM charge of the Hawking quanta plus the charge of the nucleon remnant.
If the EM charge is conserved and $n_p=0$, the BH remnant is assumed to be
electrically neutral, i.e.\ $Q_R=0$: The absolute value of the total charge in
the Hawking quanta is $|Q_{H}|\leq 2e/3$ and $Q_N=1-Q_H$. This is justified from
the fact that the BH charge should have been shed earlier in the evaporation
process. (See, however, Ref.\ \cite{Koch:2005ks} for a different viewpoint.) It
should be stressed that the air shower phenomenology of a charge remnant is not
known and it is not clear how to track it in the atmosphere in a meaningful
way. 
\section{Basics of neutrino air showers}
This section presents the essentials of the theory and phenomenology of UHECR
neutrino air showers.
\subsection{Physics of neutrino air showers}
UHECRs are believed to be a composite of protons and heavier nuclei.
Ultrahigh-energy neutrinos are created as these UHECRs interact with the cosmic
microwave background through photopion production (protons) or the infrared
background (iron nuclei). A cutoff in the energy spectrum is expected at the
threshold energy of the photopion production, known as the
Greisen-Zatsepin-Kuzmin (GZK) cutoff. These neutrinos are called cosmogenic or
GZK neutrinos. Cosmogenic neutrinos are almost ``guaranteed'' to exist, though
they have not been observed yet. They are the most likely source of neutrinos
to produce BHs in the atmosphere. The peak of the cosmogenic neutrino flux is
around $10^{17 -18}$ eV (c.m.\ energy $\sim$ 10 -- 50 TeV). The flux depends on
a number of parameters of the UHECR source such as spatial distribution,
injection spectrum, abundance, maximum energy and cosmological evolution. These
factors can affect the flux even by a couple of orders of magnitude.

The depth of the first interaction point, $X_0$, depends on the total cross
section of the process. The column depth of Earth's atmosphere in the
horizontal direction is $3.6 \times 10^4$ g~cm$^{-2}$. The interaction length
of a neutrino with energy $E_\nu=10^{9}$ TeV is $\lambda_{CC} \simeq 1.1 \times
10^7$ g~cm$^{-2}$ for charged current (CC) interactions. The largest possible
cross sections of BH events give shorter interaction lengths, but still larger
than the column depth of Earth's atmosphere. Therefore, neutrinos interacting
in the atmosphere can induce air showers at any $X_0$. In contrast, SM hadronic
interactions have large cross sections with $X_0$ high in the atmosphere.
Considering deeply penetrating horizontal air showers effectively filters out
SM hadronic air showers, while giving the most likely chance of interaction.
The background for detecting BH air showers is limited to SM neutrino air
showers.

The SM interaction channels are the CC and neutral current (NC) for all three
flavors. The energy of the leading lepton in the final state is given by $(1-y)
E_\nu$, where the mean value of the inelasticity $y$ is $\approx 0.2$. The
leading lepton in the NC interaction is a neutrino that does not contribute to
the air shower and the cross section is lower than the cross section of the CC
interaction, $\sigma_{NC}(E) \approx 0.4 \, \sigma_{CC}(E)$. The $\nu_\mu$-CC
produces a high energy $\mu$ that does not decay before reaching ground. The
$\nu_\tau$-CC produces a $\tau$ that also generally does not decay before
reaching ground. Therefore, the most relevant background for BH air showers is
the $\nu_e$-CC channel. 
\subsection{Air shower detection}
Extensive air showers can be detected with fluorescence telescopes and ground
arrays. Fluorescence telescopes observe the fluorescence light produced by the
interaction of atmospheric nitrogen molecules with the EM component of the
developing air shower. The fluorescence method pioneered by the Fly's Eye
detector \cite{flyseye} and currently operated by HiRes \cite{hires} and the
Pierre Auger Observatory (PAO) \cite{pao} is able to reconstruct the
longitudinal development of the (mainly) $e^+e^-$ component of the air shower.
This technique provides a good estimate of the energy of the primary particle
that initiates the air shower, since most of the energy of the air shower goes
into the observable EM channel. This method also enables reconstruction of the
shower maximum $X_m$, i.e.\ the depth at which the cascade contains the maximum
number of $e^+e^-$ pairs. The quantity $X_m-X_0$ is sensitive to the type of
primary particle, its energy, and the kind of interaction initiating the
cascade. The duty cycle is approximately 10\%, as clear moonless nights are
required.

Ground arrays record the ``footprint'' of the air shower. Various methods are
used to detect charged particles on the ground. Some examples are plastic
scintillators \cite{volcanoranch, agasa} and water Cerenkov tanks
\cite{haverahpark, pao}. These detectors are spread over areas ranging from a
few km$^2$ to a few thousand km$^2$. For example, the fully functioning
southern PAO will have 1600 water Cerenkov tanks each with surface area of 10
m$^2$, covering 3000 km$^2$. Arrival time, composition, and pattern of the
ground signals are used to analyze the properties of the air shower. Ground
arrays can be operated full time.

The best method of BH air shower detection is a combination of fluorescence and
ground detectors, such as the PAO. In view of this, we study both the air
shower longitudinal development and the muon content at ground level. For
inclined air showers, the geomagnetic field affects the distribution of
particles on the ground, which is very sensitive to the zenith angle.
Therefore, we simply count the number of particles rather than study their
distribution.
\section{Air shower generator}
The BH generator output consists of a list of elementary SM quanta which are
decayed with PYTHIA. The secondaries of the BH decay (PYTHIA output) are
boosted to the laboratory frame and injected in the air shower generator AIRES
to obtain the air shower. In this section we describe in detail the air shower
part of the simulation.

The AIRES code requires to input primary cosmic ray properties and simulation
conditions. The required physical parameters are:
\begin{enumerate}
\item Energy of the primary cosmic ray;
\item Zenith angle of the primary cosmic ray;
\item Azimuth angle of the primary cosmic ray;
\item Total number of air shower simulations;
\item Starting point of the air showers;
\item Number of observing levels for the longitudinal air shower development;
\item Observation site to determine geomagnetic field and ground altitude;
\item Thinning level;
\item Thinning weight limitation;
\item Threshold energies for gamma rays, electrons, muons, mesons, and nucleon;
\item Threshold energies for (approximately) propagating gamma rays and
electrons.
\end{enumerate}
In our simulations, the zenith angle is set to $70^\circ$ and $X_0$ is set to
an altitude of 10 km, corresponding to a slant depth of 780 g~cm$^{-2}$. The BH
interaction takes place at the injection altitude. The threshold energies for
tracking particles in the air showers are 100 keV for gamma rays, electrons and
positrons, 1 MeV for muons, 1.5 MeV for mesons, and 150 MeV for nucleons. The
geomagnetic field is set to the Pierre Auger Observatory (El Nihuil site). The
thinning level is $10^{-6}$ with weight limitation of 0.2. A more detailed
explanation of each parameter and other possible options can be found in the
AIRES manual. 
\subsection{BH air showers}
The steps to initiate a BH air shower are:
\begin{enumerate}
\item The BH is decayed in the c.m.\ frame. The unstable quanta are hadronized
or decayed instantaneously by PYTHIA, with the exception of top quarks and
$\tau$ leptons. PYTHIA does not handle top quarks. Therefore, they are
instantaneously decayed as $t \to bW$ before being injected in PYTHIA. The
$\tau$ leptons are produced directly from BH evaporation, or from hadronization
or decay of other particles. Depending on their energy, the $\tau$s may decay
before reaching the ground. In that case, they are decayed with PYTHIA but
their secondaries are injected into AIRES at different atmospheric depths,
according to their boost and free path (see below). 
\item All secondaries from PYTHIA are boosted to the laboratory frame. The
particles are tightly beamed due to their very high boost. 
\item All secondaries are injected into AIRES as primaries of the air shower.
\end{enumerate}
Note that neutral pions generated in the hadronization process are immediately
decayed by PYTHIA in the c.m.\ frame. Their average energy in the laboratory
frame is smaller than the critical energy, making them more likely to decay
than interact.
\subsection{SM air showers}
The CC and NC SM air shower simulations follow Ref.~\cite{Ahn:2003qn}:
\begin{enumerate}
\item The differential cross section is integrated over the fraction of the
total nucleon momentum carried by the parton for all possible values of $y$.
\item $y$ is sampled from the previous distribution. The energy of the leading
lepton is $(1-y) E_\nu$.
\item The leading lepton of the CC interaction is injected into AIRES. The
leading neutrino of the NC interaction is not observable and is not injected.
\item The hadronic part of CC and NC interactions are hadronized with PYTHIA in
the c.m.\ frame. The resulting particles are boosted back into the laboratory
frame and injected into AIRES.
\end{enumerate}
The $\tau$ lepton in the $\nu_\tau$-CC interaction is treated separately as in
the BH air showers. The $\tau$ is decayed with PYTHIA and its secondaries are
injected at the corresponding atmospheric depth.
\section{Simulation results: SM vs.\ BH air shower detection in fluorescence
detectors \label{simBHSMfl}}
Simulations show that the characteristics of BH air showers and SM $\nu_e$-CC
air showers with identical first interaction point $X_0$ are quite different.
However, these differences can vanish with a suitable shift of $X_0$ for either
of the two processes. As $X_0$ is not a fixed parameter for these interactions,
the properties of BH air showers are of limited practical use for detection
purposes. The characteristics of BH air showers and SM $\nu_e$-CC air showers
are summarized in the following table:
\begin{center}
\begin{tabular}{|l||c|c|}
\hline
&~BH air showers~&~SM $\nu_e$-CC air showers~\\
\hline
~Muon content & High & Low \\
~Development & Quick & Slow \\
~Peak fluctuations & Small & Large \\
~Average total energy & Varying & Stable \\  
\hline
\end{tabular}
\end{center}
The BH air showers are similar to hadronic air showers. The $\nu_e$-CC air
showers are comparable to air showers generated by photons. The hadronic nature
of the BH air showers is due to the prevalence of hadronic channels in the
Hawking evaporation phase. Their rapidity is due to the large number of hadrons
initiating the air showers. On the contrary, the main interaction channels in
the SM events are pair production and bremsstrahlung. These processes produce a
smaller number of secondaries than a hadronic interaction. The
Landau-Pomeranchuk-Migdal effect \cite{lpm} also contributes in slowing the air
shower for primary energies $\gtrsim 10^7$ TeV \cite{aharonian}. The BH air
showers exhibit smaller fluctuations in $X_m$ than the $\nu_e$-CC air showers.
Although the BH mass varies from shower to shower, their development is more or
less stable because of the large number of BH secondaries. The BH air showers
can be viewed as a superposition of many air showers with less energy. The
larger fluctuations from shower to shower in the SM process are due to the
fluctuations in the energy $(1-y)E_\nu$ carried by the leading lepton. The
large variations in the total energy of the BH air showers are due to the
presence of invisible channels (gravitons, neutrinos, and non-decaying
$\tau$s). For example, the proton remnant may fragment into a top quark, which
decays as $t\to b+W$. If the $W$ decay mode is leptonic, a consistent part of
the initial proton remnant energy may be carried away by a neutrino. On the
contrary, most of the c.m.\ energy in the $\nu_e$-CC interaction is observable.
This leads to a stable air shower total energy.

The differences between SM and BH air showers can be quantified by choosing a
benchmark model for the BH process and comparing this model to the SM process.
This method also allows to differentiate the effects of various parameters and
theoretical models from the stable characteristics of the BH air showers. With
some guidance from the theory, a reasonable choice is:
\begin{center}
\begin{tabular}{|l||c|}
\hline
~Parameter~&~Benchmark value ~\\
\hline
~Planck mass ($M_\star$) & 1 TeV\\
~Number of extra dimensions ($n$) & 6\\
~Formation model& Black disk\\
~Minimum-allowed mass ($M_{min}$)& $2~M_\star$\\
~Quantum threshold ($Q_{min}$) & $1~M_\star$\\
~Final hard quanta ($n_p$) & 2\\
~Momentum transfer ($Q$) & $R^{-1}(M)$\\
~EM charge conservation & YES\\
~Minimum spacetime length ($l_m$)~& 0\\
\hline
\end{tabular}
\end{center}
\begin{figure}
\includegraphics[width=0.8\textwidth]{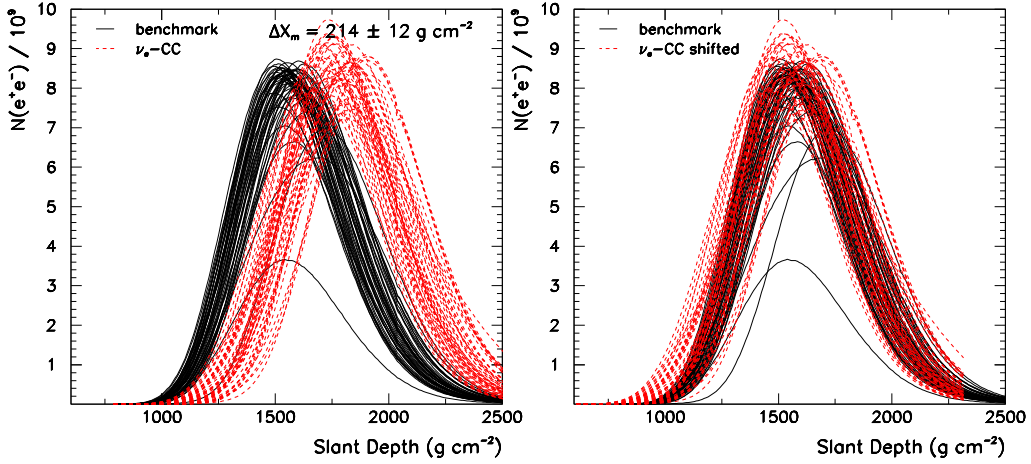}
\caption{Number of $e^+e^-$ vs.\ slant depth for the longitudinal development
of 50 air showers with $E_\nu = 10^7$ TeV. The BH air showers for the benchmark
model (black solid lines) and the $\nu_e$-CC air showers (red dashed lines) are
shown. The air shower maxima are $X_m = 1566 \pm 6$ g~cm$^{-2}$ for the BH
benchmark model and $X_m = 1780 \pm 9$ g~cm$^{-2}$ for the $\nu_e$-CC,
respectively. The difference in the air shower maxima is $\Delta X_m=214 \pm
12$ g~cm$^{-2}$. The left panel has both air showers with identical first
interaction point, $X_0(\nu_e$-CC)\,$= X_0(BH)$ = 780 g~cm$^{-2}$. The right
panel shows the same air showers with a shift in $X_0(\nu_e$-CC) such that
$X_m(\nu_e$-CC)\,$\simeq X_m(BH)$.}
\label{bhnueccfig}
\end{figure}
The left panel of Fig.~\ref{bhnueccfig} shows the $\nu_e$-CC air showers and
the benchmark BH air showers (50 runs each, neutrino primary energy $E_\nu=
10^7$ TeV). The difference in the shower maxima is $\Delta X_m=214 \pm 12$
g~cm$^{-2}$. Although the showers appear to be quite distinct, this difference
is a consequence of the same choice of $X_0$ for both BH and SM air showers.
The right panel of the figure shows $\nu_e$-CC air showers shifted so $X_m(BH)
\simeq X_m(\nu_e$-CC). Therefore, BH and SM air showers can only be
distinguished when $X_0- X_m$ is clearly measured. Since present detectors
cannot measure $X_0$, $\nu_e$-CC air showers and BH air showers cannot be
discriminated on an event-by-event basis \cite{Ahn:2003qn}. We will see in
Sect.\ \ref{simparam} that this conclusion does not substantially change if the
BH parameters are varied.
\section{Simulation results: SM vs.\ BH air shower detection in ground and
hybrid detectors \label{simBHSMhb}}
\begin{figure}
\includegraphics[width=0.6\textwidth]{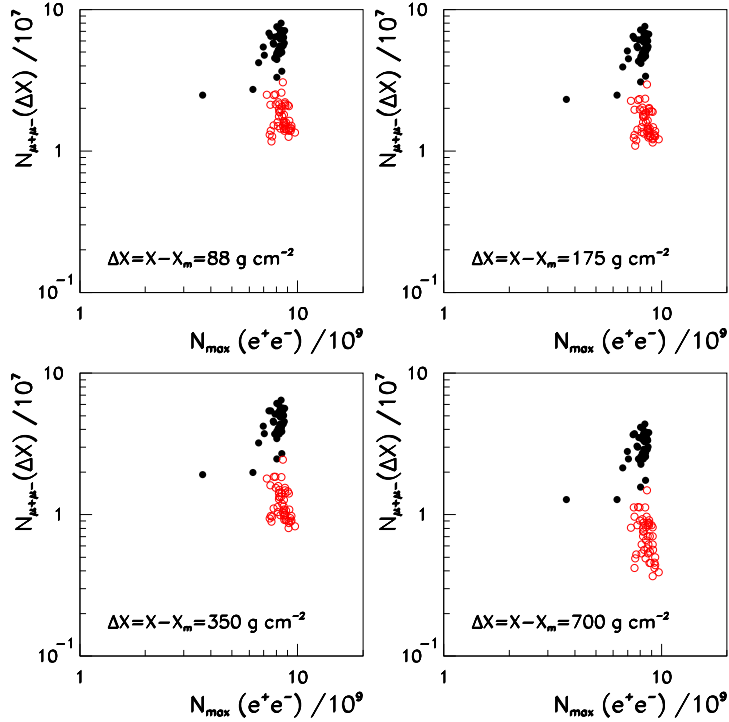}
\caption{Number of $\mu^+\mu^-$ at various atmospheric depths $X_m + \Delta X$
vs.\ the number of $e^+e^-$ at $X_m$ for 50 benchmark model BH air showers
(black filled circles) and 50 $\nu_e$-CC air showers (red empty circles). The
energy of the primary neutrino is $E_\nu = 10^7$ TeV. The observation depth of
the muons increases from left to right panel and from top to bottom panel.}
\label{nmubh}
\end{figure}

Discrimination of BH air showers and $\nu_e$-CC air showers can be improved by
the use of ground arrays. The best possible scenario for BH air shower
detection is a technique that combines air fluorescence telescopes and a ground
array. Since fluorescence telescopes are able to measure $X_m$ accurately, a
good air shower discriminator is to fix $X_0$ and count the number of particles
at various distances from $X_m$. This is equivalent to fixing the detection
level and varying the air shower first interaction point. The fluctuation due
to the change of $X_0$ is negligible compared to fluctuations arising from other
uncertainties. Figure~\ref{nmubh} shows the number of muons at different
atmospheric depths vs.\ the number of electrons at the air shower maximum for
BH air showers (benchmark model) and $\nu_e$-CC air showers. The muons are
measured from the ground array and the electrons are measured from the
fluorescence telescopes.
The BH air showers are characterized by a higher muon content than the
$\nu_e$-CC air showers. Although the separation is not large enough to
distinguish the air showers on an event-by-event basis, discrimination of
BH and SM events is possible with enough statistics. The number of muons also
depends on the detection level. Since different $\Delta X$s can be seen as
different initial interaction points, the air showers in the lower right (upper
left) panel of Fig.~\ref{nmubh} can be understood as starting higher (lower) in
the atmosphere than the air showers in the other panels. The number of muons in
the air showers decreases if the first interaction point is higher in the
atmosphere. 

\begin{figure}
\includegraphics[width=0.6\textwidth]{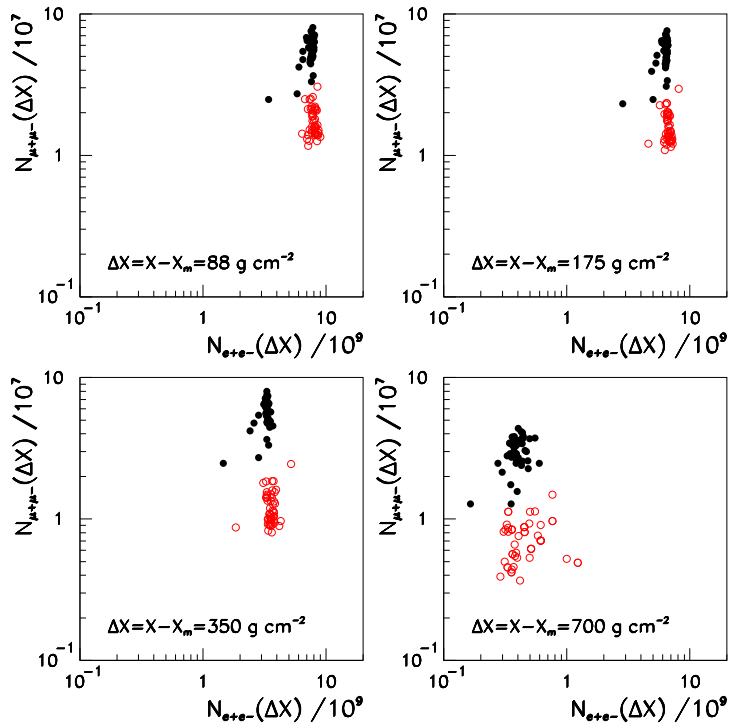}
\caption{Number of $\mu^+\mu^-$ vs.\ number of $e^+e^-$at various depths $X_m +
\Delta X$. The comparison is between 50 BH benchmark model air showers (black
filled circles) and 50 $\nu_e$-CC air showers (red empty circles) with $E_\nu =
10^7$ TeV.The observation depth of the muons and electrons increases from left
to right panel and from top to bottom panel.}
\label{emmubh}
\end{figure}

If only ground detection is possible, a good BH vs.\ SM discriminator is the
number of electrons and muons at various atmospheric depths $X_m + \Delta X$
(Fig.\ \ref{emmubh}). As in the hybrid detection scenario, although the BH air
showers show higher muon content than the $\nu_e$-CC air showers, their
discrimination requires large statistics. The number of muons depend on the
observation level and decreases as $\Delta X$ increases. In absence of an air
fluorescence telescope to accurately measure $X_m$, alternative techniques must
be used to reconstruct the air shower maximum.
\section{Simulation results: Effect of BH parameters \label{simparam}}
\begin{figure}
\includegraphics[width=0.8\textwidth]{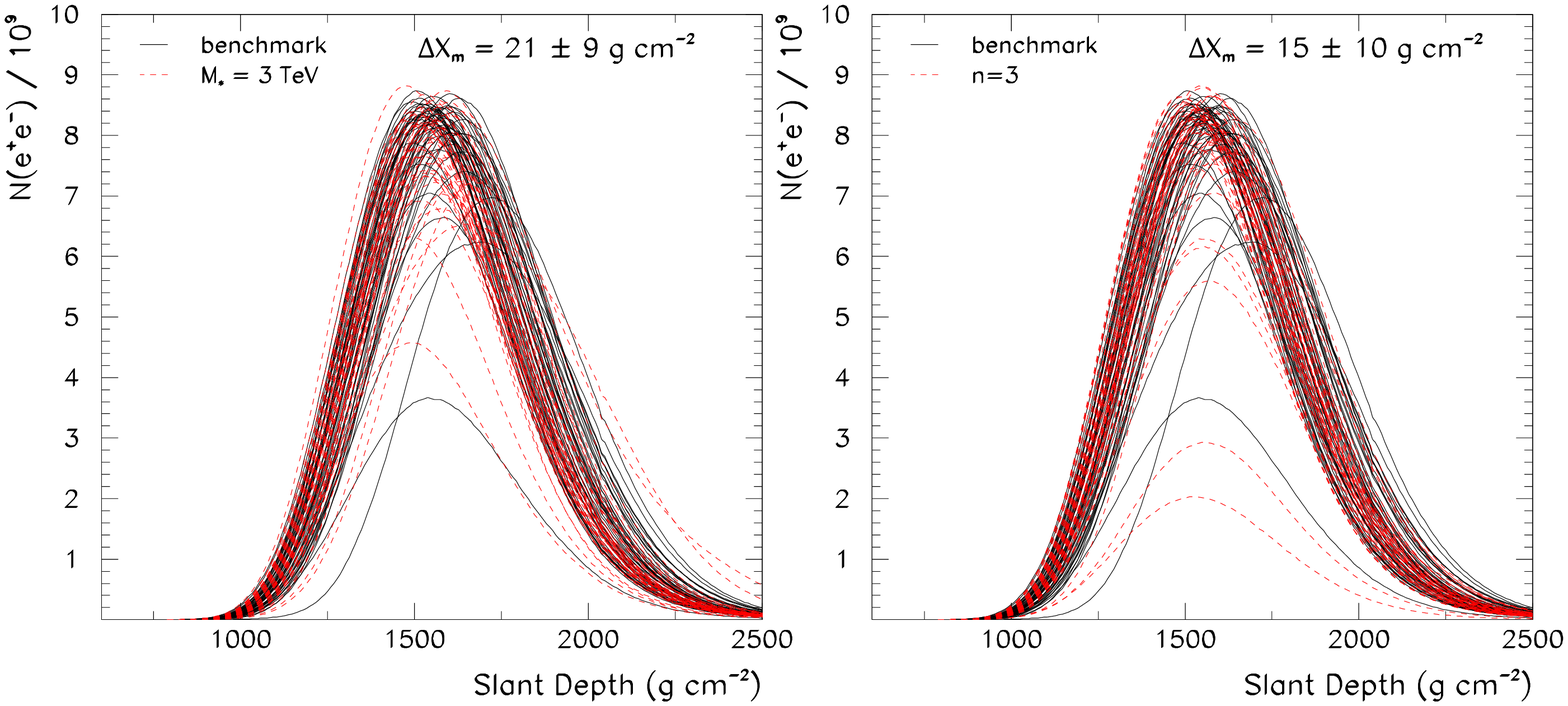}
\caption{Longitudinal development of 50 air showers for the BH benchmark model
(black solid curves) vs.\ two different choices of BH parameters (red dashed
curves). The energy of the primary neutrino is $E_\nu = 10^7$ TeV. The left
panel shows the difference between the benchmark model ($M_\star=1$ TeV) and
$M_\star = 3$ TeV. The average difference in the air shower maxima is $\Delta
X_m=21 \pm 9$ g~cm$^{-2}$. The right panel shows the benchmark case ($n=6$)
and $n=3$. The average difference in $X_m$ is $\Delta X_m = 15 \pm 10$
g~cm$^{-2}$.}
\label{bhmstn}
\end{figure}

\begin{figure}
\includegraphics[width=0.8\textwidth]{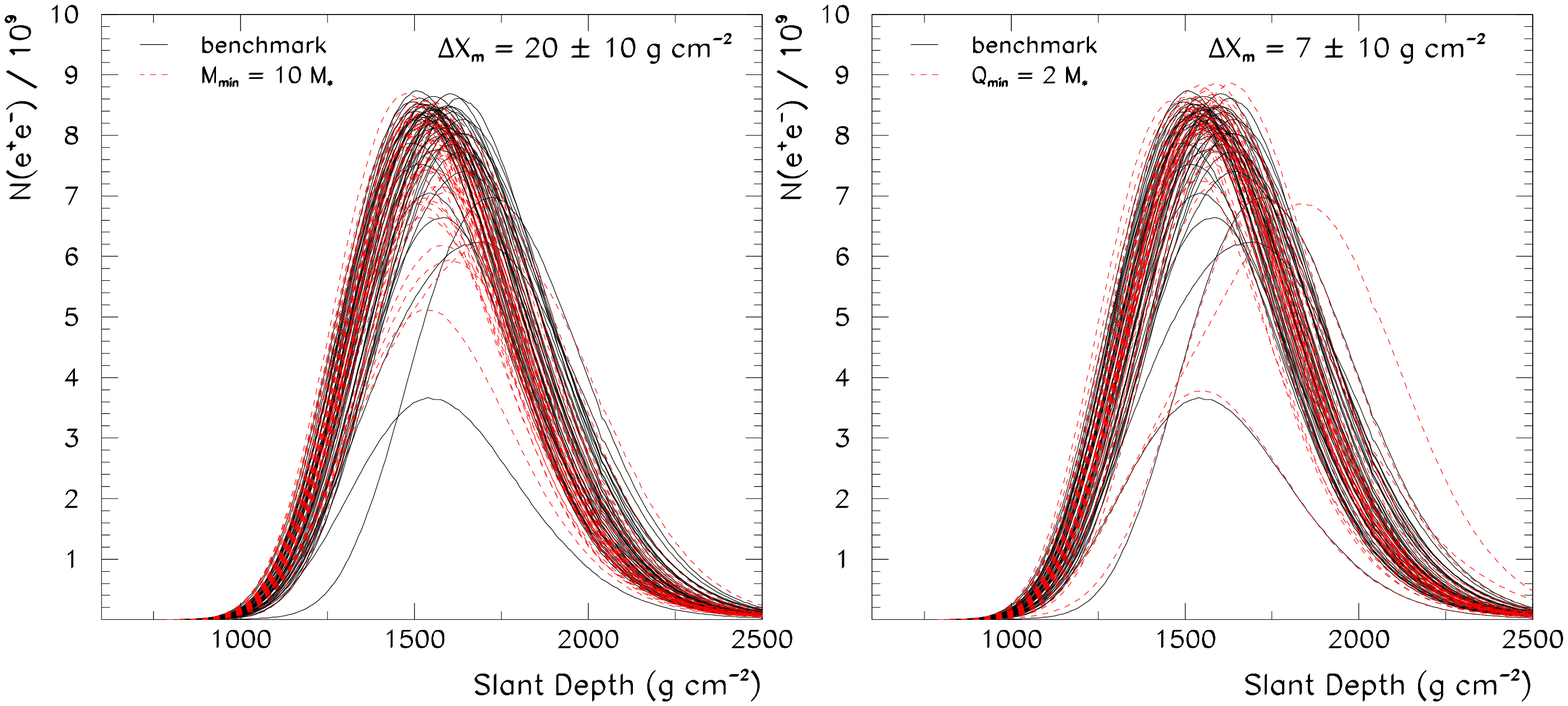}
\caption{Longitudinal development of 50 air showers for the BH benchmark model
(black solid curves) vs.\ two different choices of BH parameters (red dashed
curves). The energy of the primary neutrino is $E_\nu = 10^7$ TeV. The left
panel shows the difference between the benchmark model ($M_{min}=2M_\star$) and
$M_{min} = 10M_\star$. The average difference in the air shower maxima is
$\Delta X_m=20 \pm 10$ g~cm$^{-2}$. The right panel shows the benchmark case
($Q_{min}=M_\star$) and $Q_{min} = 2M_\star$. The average difference in $X_m$
is $\Delta X_m = 7 \pm 10$ g~cm$^{-2}$. }
\label{bhnueccfig2}
\end{figure}

\begin{figure}
\includegraphics[width=0.8\textwidth]{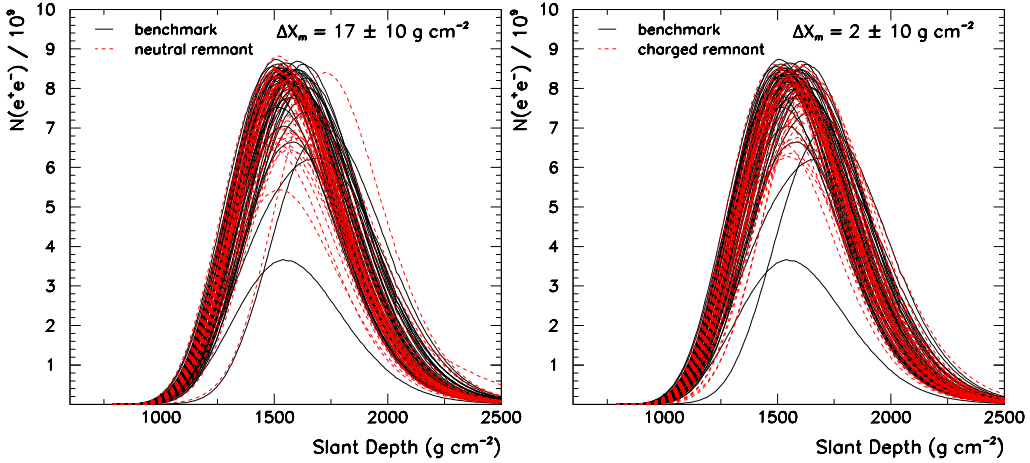}
\caption{Longitudinal development of 50 air showers for the BH benchmark model
(black solid curves) vs.\ two different choices of BH parameters (red dashed
curves). The energy of the primary neutrino is $E_\nu = 10^7$ TeV. The left
panel shows the difference between the benchmark model (final decay in 2
quanta) and BH evolution with final electrically neutral remnant. The average
difference in the air shower maxima is $\Delta X_m=17 \pm 10$ g~cm$^{-2}$. The
right panel shows the benchmark case and  BH evolution with final electrically
charged remnant. The average difference in $X_m$ is $\Delta X_m = 2 \pm 10$
g~cm$^{-2}$.}
\label{bhnremcrem}
\end{figure}

\begin{figure}
\includegraphics[width=0.8\textwidth]{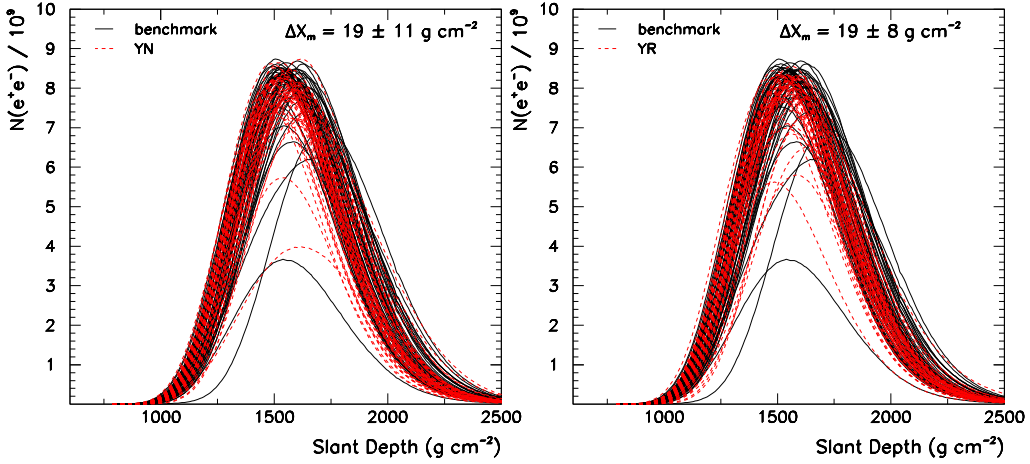}
\caption{Longitudinal development of 50 air showers for the BH benchmark model
(black solid curves) vs.\ two different choices of BH parameters (red dashed
curves). The energy of the primary neutrino is $E_\nu = 10^7$ TeV. The left
panel shows the difference between the benchmark model (black disk) and the YN
graviton loss model. The average difference in the air shower maxima is $\Delta
X_m=19 \pm 11$ g~cm$^{-2}$. The right panel shows the benchmark case and the
improved YR graviton loss model. The average difference in $X_m$ is $\Delta X_m
= 19 \pm 8$ g~cm$^{-2}$. There is virtually no difference between the YN and
YR models.}
\label{bhynyr}
\end{figure}

\begin{figure}
\includegraphics[width=0.4\textwidth]{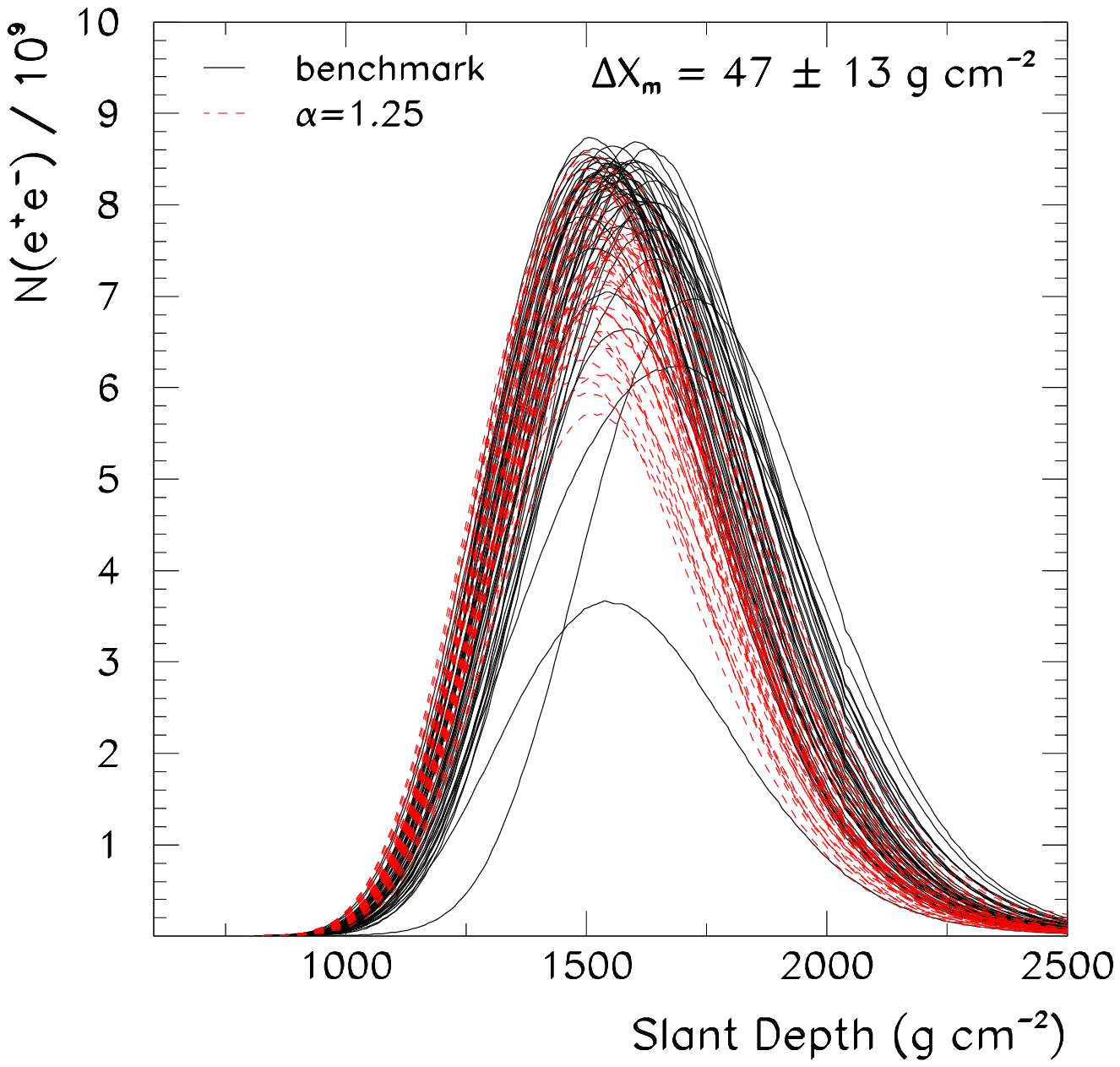}
\caption{Longitudinal development of 50 air showers for the BH benchmark model
(black solid curves) and a BH evolution model with minimum length $l_{min} =
2\alpha ~M_\star^{-1}$  (red dashed curves). The energy of the primary neutrino
is $E_\nu = 10^7$ TeV. The average difference in the air shower maxima is
$\Delta X_m=47 \pm 13$ g~cm$^{-2}$.}
\label{bhalpha}
\end{figure}

The effects of various BH models can be studied by varying the parameters
described in Sect.\ \ref{BHgenerator} and comparing the simulations to the
benchmark model. The air shower longitudinal developments for different choices
are shown in Figs.\ \ref{bhmstn}-\ref{bhalpha} and summarized in Table II.

\begin{table}[ht]
\begin{center}
\begin{tabular}{|l|c|}
\hline
~Difference from benchmark~ & ~$X_m \rm \,\rm$ $\pm$ rms error~ \\
\hline
~None (benchmark) & $1566 \,\pm\, 6$ \\
~$M_\star = 3$ TeV & $1545 \,\pm\, 6$ \\
~$n=3$ & $1551 \,\pm\, 5$ \\
~$M_{min} = 10$ TeV & $1546 \,\pm\, 6$ \\
~$Q_{min} = 2$ TeV & $1559 \,\pm\, 7$ \\
~Neutral remnant & $1549 \,\pm\, 6$ \\
~Charged remnant & $1564 \,\pm\, 6$ \\
~YN model & $1547 \,\pm\, 6$ \\
~YR model & $1547 \,\pm\, 5$ \\
~$l_{min} = 2.5 \, M_\star^{-1}$ & $1519 \,\pm\, 4$ \\
\hline
\end{tabular}
\caption{Shower maximum $X_m$ and rms error for 50 BH air showers with
different physical parameters and models. The primary neutrino energy is $E_\nu
= 10^7$ TeV. The first row gives $X_m$ for the benchmark model of Sect.\
\ref{simBHSMfl}.}
\end{center}
\label{bhvary}
\end{table}

The depth of the air shower maximum is not significantly affected by changes in
the BH parameters or the theoretical model. Fluctuations in the air showers are
large; all $X_m$ values lie within one standard deviation from each other, with
the possible exception of the simulations with nonzero minimum length. We
conclude that the main characteristics of BH air showers described in the
previous section are robust. This result can be qualitatively explained by
noticing that most of the c.m.\ energy of the neutrino-nucleon collision is not
trapped in the BH. Therefore, different choices of BH parameters do not produce
large observable effects in the air shower development. The main factors
determining the BH evolution (BH mass distribution, energy and spectrum of
emitted quanta) are difficult to disentangle because their variations do not
act to coherently increase or decrease the shower maximum. For instance,
increasing the minimum-allowed BH mass from $M_{min}=2$ TeV to $M_{min}=10$ TeV
increases the average BH mass in the air showers. This leads to a larger number
of quanta. However, this property does not translate into a faster air shower
development; the average energy per quanta is smaller, and the two effects
compensate each other. Two interesting facts are worth observing. Firstly, the
benchmark case has the largest cross section and the largest $X_m$. This result
is mainly due to the choice of a relatively small fundamental Planck constant.
Adding quantum effects or graviton loss at formation seems to decrease $X_m$
slightly. Secondly, the presence of a minimum length may possibly be the only
BH physical signature distinguishable from the black disk model. However, in
our simulations the choice of $l_{min}$ has been purposedly fine-tuned to the
maximum-allowed value that allows BH formation with primary neutrino energy
$E_\nu=10^7$ TeV. Relaxing this choice leads to values of $X_m$ closer to the
black disk result.   
\section{Simulation results: Background-free signal from $\tau$ decay}
It has been suggested that $\tau$ leptons produced during BH events, either
directly or through decay of other particles, can produce an observable signal
similar to the double bang produced in $\nu_\tau$-CC interactions
\cite{Cardoso:2004zi}. The mechanism is the following. The BH evaporation
initiates a first air shower, while a second air shower is initiated by the
$\tau$ decay at a lower altitude. If the second bang is large enough to be
observed, the double peak feature provides a background-free signature
independent of  the first interaction point. 

The $\tau$ lepton in the $\nu_\tau$-CC interaction carries on average about
80\,\% of the total c.m.\ energy. In a BH event, the fraction of the total BH
mass going into the $\tau$ second bang can be estimated from Eq.~(\ref{multii})
to be roughly
\begin{equation}
\epsilon_\tau \,\sim\, \frac{2}{\sum_j c_j\Gamma_{j}f_j(3)}\sim 4\%\,.
\label{epsilontau}
\end{equation}
Therefore, a $\tau$ from BH evaporation carries on average a few percent of the
total BH mass. Since the energy trapped in the BH accounts generally only for a
small fraction of the initial c.m.\ energy, the total energy in the $\tau$
channel is on average less than 1\% of the total air shower energy. This gives
the $\tau$ a higher chance to decay before reaching ground, but a smaller
energy deposit in the second bang, making the latter harder to detect. 
Moreover, $\tau$s are not produced every time a BH is formed and the
probability to observe an air shower with a $\tau$ decaying before reaching
ground is relatively small.

\begin{figure}
\includegraphics[width=0.8\textwidth]{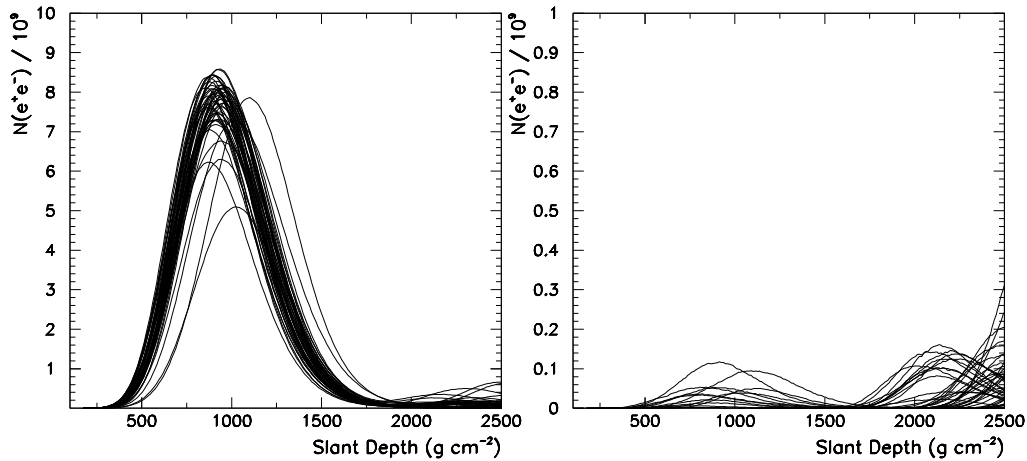}
\caption{Longitudinal development of 50 BH air showers containing at least one
$\tau$ decaying at an altitude $X> 0.75(X_g-X_0)$. The air showers are started
at an altitude of 20 km, corresponding to a slant depth of $X_0 = 160$
g~cm$^{-2}$. The left panel shows the profile of all the particles in the air
shower. The right panel shows only the profile of the secondary particles from
the $\tau$ decay. Note that the y axis of the right panel is magnified 10 times
w.r.t\ the left panel. The peaks between 500 g~cm$^{-2}$ and 1500 g~cm$^{-2}$
are originated by multiple $\tau$ events with one $\tau$ decaying high in the
atmosphere.}
\label{taufig}
\end{figure}

The double bang signature can be studied by selecting only air showers
containing at least one $\tau$ decaying in air. We simulated these air showers
using the benchmark parameters, zenith angle $70^\circ$ and $X_0=160$
g~cm$^{-2}$ (altitude of the first interaction point = 20 km). The higher
altitude gives a larger separation between the two bangs. Two possible
scenarios were considered: i) $\tau$s decaying at any depth in the atmosphere,
and ii) at least one $\tau$ decaying at an altitude $X> 0.75(X_g-X_0)$, where
$X_g$ is the slant depth of the ground (``low altitude $\tau$s''). Case ii)
represents the best possible scenario for double bang detection, as the second
bang is expected to occur close to the detector. The longitudinal profiles of
the low altitude $\tau$ air showers are shown in the left panel of
Fig.~\ref{taufig}. The second bang is visible around 2250 g~cm$^{-2}$. The
longitudinal development of the $\tau$ channel contribution to the air shower
is plotted in the right panel. The peaks between 500 g~cm$^{-2}$ and 1500
g~cm$^{-2}$ are from events with multiple $\tau$s, where at least one of these
$\tau$s decays at low altitude. As is expected, the $\tau$ component
contributes only a minimal fraction to the overall air shower energy. Although
the second bang is in principle detectable, it is within the fluctuation of the
first bang's tail; current and next generation detectors cannot discern a few
percent feature. Moreover, existing detectors such as the PAO have a limited
field of view and cannot track the full profile.
\section{\label{concl} Conclusions}
We simulated extensive air showers initiated by TeV-scale BH events produced
from neutrino interaction in Earth's atmosphere. These simulations were
performed with an advanced fortran MC code \cite{Groke1.01:2005} that includes
most of the theoretical results of the recent literature. The BH air showers
were compared to SM air showers and different models of BH formation and
evolution were investigated. We also studied exotic signatures of BH events,
such as the ``$\tau$ double bang'' signature. Our goal was to test various
proposals of BH vs.\ SM air shower discrimination and look for new ways of
differentiating models of BH formation and evolution.  

Our results show that the main features of BH air showers are largely
independent of the details of BH formation and evolution. Statistical
fluctuations and limitations in detection techniques hinder the discrimination
of alternative theoretical or phenomenological models. No difference between
the black disk model and alternative models of BH formation, or between BHs
with final explosive decay and stable remnant, can be detected with current
UHECR observatories. Distinguishing SM and BH air showers with hybrid detectors
is possible if enough statistics is gathered. The most promising way is to
measure the air shower maximum with a fluorescence telescope and count muons at
ground.  The double bang signature cannot be observed by any realistic detector
at the present stage. These results imply that the theory of TeV-scale BHs in
UHECRs is robust, but BH air showers are hardly to probe details of ``new
physics'' in the near future.
\section*{Acknowledgements}
We are grateful to D.\ Allard, M.\ Ave, N.\ Busca, V.\ Cardoso, A.V.\ Olinto
and H.P.\ de Bretagne for discussions and many useful suggestions. We warmly
thank H.\ Yoshino for providing the numerical tables of the apparent horizon
mass for the trapped-surface models.  This research was carried out at the
University of Chicago, Kavli Institute for Cosmological Physics and at the
University of Mississippi. It was supported (in part) by grant NSF PHY-0114422
and a University of Mississippi FRP grant. KICP is a NSF Physics Frontier
Center.


\begin{thebibliography}{99}

\bibitem{Amati:1987wq}
  D.~Amati, M.~Ciafaloni and G.~Veneziano,
  Phys.\ Lett.\ B {\bf 197}, 81 (1987);
  D.~Amati, M.~Ciafaloni and G.~Veneziano,
  Int.\ J.\ Mod.\ Phys.\ A {\bf 3}, 1615 (1988);
  H.~Verlinde and E.~Verlinde,
  Nucl.\ Phys.\ B {\bf 371}, 246 (1992)
  [arXiv:hep-th/9110017].


\bibitem{Banks:1999gd}
  T.~Banks and W.~Fischler,
  arXiv:hep-th/9906038.


\bibitem{Arkani-Hamed:1998rs}
  N.~Arkani-Hamed, S.~Dimopoulos and G.~R.~Dvali,
  Phys.\ Lett.\ B {\bf 429}, 263 (1998)
  [arXiv:hep-ph/9803315];
  I.~Antoniadis, N.~Arkani-Hamed, S.~Dimopoulos and G.~R.~Dvali,
  Phys.\ Lett.\ B {\bf 436}, 257 (1998)
  [arXiv:hep-ph/9804398];
  N.~Arkani-Hamed, S.~Dimopoulos and G.~R.~Dvali,
  Phys.\ Rev.\ D {\bf 59}, 086004 (1999)
  [arXiv:hep-ph/9807344].


\bibitem{Maartens:2003tw}
  R.~Maartens,
  Living Rev.\ Rel.\  {\bf 7}, 7 (2004)
  [arXiv:gr-qc/0312059].


\bibitem{Giddings:2001bu}
  S.~B.~Giddings and S.~Thomas,
  Phys.\ Rev.\ D {\bf 65}, 056010 (2002)
  [arXiv:hep-ph/0106219];
  S.~Dimopoulos and G.~Landsberg,
  Phys.\ Rev.\ Lett.\  {\bf 87}, 161602 (2001)
  [arXiv:hep-ph/0106295];
  K.~m.~Cheung,
  Phys.\ Rev.\ Lett.\  {\bf 88}, 221602 (2002)
  [arXiv:hep-ph/0110163];
  A.~Chamblin and G.~C.~Nayak,
  Phys.\ Rev.\ D {\bf 66}, 091901 (2002)
  [arXiv:hep-ph/0206060];
  A.~Chamblin, F.~Cooper and G.~C.~Nayak,
  Phys.\ Rev.\ D {\bf 70}, 075018 (2004)
  [arXiv:hep-ph/0405054];
  R.~Godang, S.~Bracker, M.~Cavagli\`a, L.~Cremaldi, D.~Summers and D.~Cline,
  Int.\ J.\ Mod.\ Phys.\ A {\bf 20}, 3409 (2005)
  [arXiv:hep-ph/0411248].

\bibitem{Ahn:2002mj}
  E.~J.~Ahn, M.~Cavagli\`a and A.~V.~Olinto,
  Phys.\ Lett.\ B {\bf 551}, 1 (2003)
  [arXiv:hep-th/0201042];
  E.~J.~Ahn and M.~Cavagli\`a,
  Gen.\ Rel.\ Grav.\  {\bf 34}, 2037 (2002)
  [arXiv:hep-ph/0205168];
  K.~Cheung and C.~H.~Chou,
  Phys.\ Rev.\ D {\bf 66}, 036008 (2002)
  [arXiv:hep-ph/0205284].


\bibitem{Feng:2001ib}
  J.~L.~Feng and A.~D.~Shapere,
  Phys.\ Rev.\ Lett.\  {\bf 88}, 021303 (2002)
  [arXiv:hep-ph/0109106];
  L.~Anchordoqui and H.~Goldberg,
  Phys.\ Rev.\ D {\bf 65}, 047502 (2002)
  [arXiv:hep-ph/0109242];
  A.~Ringwald and H.~Tu,
  Phys.\ Lett.\ B {\bf 525}, 135 (2002)
  [arXiv:hep-ph/0111042];
  M.~Kowalski, A.~Ringwald and H.~Tu,
  Phys.\ Lett.\ B {\bf 529}, 1 (2002)
  [arXiv:hep-ph/0201139];
  E.~J.~Ahn, M.~Cavagli\`a and A.~V.~Olinto,
  Astropart.\ Phys.\  {\bf 22}, 377 (2005)
  [arXiv:hep-ph/0312249];
  A.~Mironov, A.~Morozov and T.~N.~Tomaras,
  arXiv:hep-ph/0311318;
  A.~Cafarella, C.~Coriano and T.~N.~Tomaras,
  JHEP {\bf 0506}, 065 (2005)
  [arXiv:hep-ph/0410358];
  J.~I.~Illana, M.~Masip and D.~Meloni,
  Phys.\ Rev.\ D {\bf 72}, 024003 (2005)
  [arXiv:hep-ph/0504234];
  M.~Ahlers, A.~Ringwald and H.~Tu,
  arXiv:astro-ph/0506698.
  

\bibitem{Anchordoqui:2001cg}
  L.~A.~Anchordoqui, J.~L.~Feng, H.~Goldberg and A.~D.~Shapere,
  Phys.\ Rev.\ D {\bf 65}, 124027 (2002)
  [arXiv:hep-ph/0112247].


\bibitem{Ahn:2003qn}
  E.~J.~Ahn, M.~Ave, M.~Cavagli\`a and A.~V.~Olinto,
  Phys.\ Rev.\ D {\bf 68}, 043004 (2003)
  [arXiv:hep-ph/0306008].


\bibitem{Cardoso:2004zi}
  V.~Cardoso, M.~C.~Esp{\`\i}rito Santo, M.~Paulos, M.~Pimenta and B.~Tom\'e,
  Astropart.\ Phys.\  {\bf 22}, 399 (2005)
  [arXiv:hep-ph/0405056].


\bibitem{Cavaglia:2002si}
  M.~Cavagli\`a,
  Int.\ J.\ Mod.\ Phys.\ A {\bf 18}, 1843 (2003)
  [arXiv:hep-ph/0210296].


\bibitem{Landsberg:2002sa}
  G.~Landsberg,
  arXiv:hep-ph/0211043;
  R.~Emparan,
  arXiv:hep-ph/0302226;
  P.~Kanti,
  Int.\ J.\ Mod.\ Phys.\ A {\bf 19}, 4899 (2004)
  [arXiv:hep-ph/0402168];
  S.~Hossenfelder,
  arXiv:hep-ph/0412265.
  
\bibitem{Cardoso:2005jq}
  V.~Cardoso, E.~Berti and M.~Cavagli\`a,
  Class.\ Quant.\ Grav.\  {\bf 22}, L61 (2005)
  [arXiv:hep-ph/0505125].


\bibitem{Hawking:1974sw}
  S.~W.~Hawking,
  Commun.\ Math.\ Phys.\  {\bf 43}, 199 (1975)
  [Erratum-ibid.\  {\bf 46}, 206 (1976)].


\bibitem{Amati:1999fv}
  D.~Amati and J.~G.~Russo,
  Phys.\ Lett.\ B {\bf 454}, 207 (1999)
  [arXiv:hep-th/9901092].
  

\bibitem{Groke1.01:2005}
  Precompiled binaries of the MC code can be freely downloaded at the web site
  http://www.phy.olemiss.edu/GR/groke.


\bibitem{Sjostrand:2000wi}
  T.~Sjostrand, P.~Eden, C.~Friberg, L.~Lonnblad, G.~Miu, S.~Mrenna and E.~Norrbin,
  Comput.\ Phys.\ Commun.\  {\bf 135}, 238 (2001)
  [arXiv:hep-ph/0010017].


\bibitem{aires}
http://www.fisica.unlp.edu.ar/auger/aires/ ~;
S.~J.~Sciutto,
arXiv:astro-ph/9905185.


\bibitem{Cavaglia:2004jw}
  M.~Cavagli\`a and S.~Das,
  Class.\ Quant.\ Grav.\  {\bf 21}, 4511 (2004)
  [arXiv:hep-th/0404050].
  
 
 \bibitem{hoop}
 K.S.~Thorne,
 in: {\it Magic without magic: John Archibald Wheeler},
 edited by J.~Klauder (Freeman, San Francisco, 1972).


\bibitem{Voloshin:2001vs}
  M.~B.~Voloshin,
  Phys.\ Lett.\ B {\bf 518}, 137 (2001)
  [arXiv:hep-ph/0107119];
  M.~B.~Voloshin,
  Phys.\ Lett.\ B {\bf 524}, 376 (2002)
  [Erratum-ibid.\ B {\bf 605}, 426 (2005)]
  [arXiv:hep-ph/0111099];
  V.~S.~Rychkov,
  Phys.\ Rev.\ D {\bf 70}, 044003 (2004)
  [arXiv:hep-ph/0401116];
  S.~B.~Giddings and V.~S.~Rychkov,
  Phys.\ Rev.\ D {\bf 70}, 104026 (2004)
  [arXiv:hep-th/0409131].

  
\bibitem{Yoshino:2002tx}
  H.~Yoshino and Y.~Nambu,
  Phys.\ Rev.\ D {\bf 67}, 024009 (2003)
  [arXiv:gr-qc/0209003].

\bibitem{Yoshino:2005hi}
  H.~Yoshino and V.~S.~Rychkov,
  Phys.\ Rev.\ D {\bf 71}, 104028 (2005)
  [arXiv:hep-th/0503171].


\bibitem{Vasilenko:2003ak}
  O.~I.~Vasilenko,
  arXiv:hep-th/0305067.


\bibitem{Aichelburg:1970dh}
  P.~C.~Aichelburg and R.~U.~Sexl,
  Gen.\ Rel.\ Grav.\  {\bf 2}, 303 (1971).


\bibitem{Kohlprath:2002yh}
  E.~Kohlprath and G.~Veneziano,
  JHEP {\bf 0206}, 057 (2002)
  [arXiv:gr-qc/0203093].

\bibitem{Cardoso:2002pa}
  V.~Cardoso, O.~J.~C.~Dias and J.~P.~S.~Lemos,
  Phys.\ Rev.\ D {\bf 67}, 064026 (2003)
  [arXiv:hep-th/0212168].


\bibitem{Berti:2003si}
  E.~Berti, M.~Cavagli\`a and L.~Gualtieri,
  Phys.\ Rev.\ D {\bf 69}, 124011 (2004)
  [arXiv:hep-th/0309203].


\bibitem{Eidelman:2004wy}
  S.~Eidelman {\it et al.}  [Particle Data Group],
  Phys.\ Lett.\ B {\bf 592}, 1 (2004).


\bibitem{Brock:1993sz}
  R.~Brock {\it et al.}  [CTEQ Collaboration],
  Rev.\ Mod.\ Phys.\  {\bf 67}, 157 (1995).
  

\bibitem{Emparan:2001kf}
R.~Emparan, M.~Masip and R.~Rattazzi,
Phys.\ Rev.\ D {\bf 65}, 064023 (2002)
[arXiv:hep-ph/0109287].


\bibitem{Yoshino:2005ps}
  H.~Yoshino, T.~Shiromizu and M.~Shibata,
  arXiv:gr-qc/0508063.


\bibitem{Anchordoqui:2003jr}
L.~A.~Anchordoqui, J.~L.~Feng, H.~Goldberg and A.~D.~Shapere,
Phys.\ Rev.\ D {\bf 68}, 104025 (2003)
[arXiv:hep-ph/0307228].
  

\bibitem{Cavaglia:2003qk}
  M.~Cavagli\`a, S.~Das and R.~Maartens,
  Class.\ Quant.\ Grav.\  {\bf 20}, L205 (2003)
  [arXiv:hep-ph/0305223].


\bibitem{Duffy:2005ns}
  G.~Duffy, C.~Harris, P.~Kanti and E.~Winstanley,
  JHEP {\bf 0509}, 049 (2005)
  [arXiv:hep-th/0507274].


\bibitem{Kanti:2002nr}
P.~Kanti and J.~March-Russell,
Phys.\ Rev.\ D {\bf 66}, 024023 (2002)
[arXiv:hep-ph/0203223];
  D.~Ida, K.~y.~Oda and S.~C.~Park,
  Phys.\ Rev.\ D {\bf 67}, 064025 (2003)
  [Erratum-ibid.\ D {\bf 69}, 049901 (2004)]
  [arXiv:hep-th/0212108];
  P.~Kanti and J.~March-Russell,
  Phys.\ Rev.\ D {\bf 67}, 104019 (2003)
  [arXiv:hep-ph/0212199];
  C.~M.~Harris and P.~Kanti,
  JHEP {\bf 0310}, 014 (2003)
  [arXiv:hep-ph/0309054];
  C.~M.~Harris and P.~Kanti,
  arXiv:hep-th/0503010;
  D.~Ida, K.~y.~Oda and S.~C.~Park,
  Phys.\ Rev.\ D {\bf 71}, 124039 (2005)
  [arXiv:hep-th/0503052];
  E.~Jung and D.~K.~Park,
  arXiv:hep-th/0506204;
  A.~S.~Cornell, W.~Naylor and M.~Sasaki,
  arXiv:hep-th/0510009.

  
\bibitem{Garay:1994en}
L.~J.~Garay,
Int.\ J.\ Mod.\ Phys.\ A {\bf 10}, 145 (1995)
[arXiv:gr-qc/9403008].


\bibitem{cteq5}
  http://durpdg.dur.ac.uk/hepdata/cteq.html


\bibitem{Cavaglia:2003hg}
  M.~Cavagli\`a,
  Phys.\ Lett.\ B {\bf 569}, 7 (2003)
  [arXiv:hep-ph/0305256].
  

\bibitem{Koch:2005ks}
  B.~Koch, M.~Bleicher and S.~Hossenfelder,
  arXiv:hep-ph/0507138.
  
 
\bibitem{flyseye}  
D.~J.~Bird {\it et al.}  [HIRES Collaboration],
Phys.\ Rev.\ Lett.\  {\bf 71}, 3401 (1993);
D.~J.~Bird {\it et al.},
Astrophys.\ J.\  {\bf 441}, 144 (1995).


\bibitem{hires}
T.~Abu-Zayyad {\it et al.},
Nucl.\ Instrum.\ Meth.\ A {\bf 450}, 253 (2000);
  T.~Abu-Zayyad {\it et al.}  [High Resolution Fly's Eye Collaboration],
  Astropart.\ Phys.\  {\bf 23}, 157 (2005)
  [arXiv:astro-ph/0208301].


\bibitem{pao}
  {\sl The Pierre Auger Project Design Report}. By Auger Collaboration.
  FERMILAB-PUB-96-024, Jan 1996. ({\sl www.auger.org}).


\bibitem{agasa}
  M.~Takeda {\it et al.},
  Phys.\ Rev.\ Lett.\  {\bf 81}, 1163 (1998)
  [arXiv:astro-ph/9807193];
  S.~Yoshida {\it et al.},
  Astropart.\ Phys.\  {\bf 3}, 105 (1995);
  N.~Hayashida {\it et al.},
  Phys.\ Rev.\ Lett.\  {\bf 73}, 3491 (1994).

\bibitem{volcanoranch}
  J.~Linsley Phys. Rev. Lett {bf\ 10}, 146 (1963);
  J.~Linsley, Proc. 8th International Cosmic Ray Conference {\bf 4}, 295 (1963)

\bibitem{haverahpark}
  M.~A.~Lawrence, R.~J.~O.~Reid and A.~A.~Watson,
  J.\ Phys.\ G {\bf 17}, 733 (1991).


\bibitem{lpm}
L.~D.~Landau and I.~Pomeranchuk,
Dokl.\ Akad.\ Nauk Ser.\ Fiz.\  {\bf 92}, 535 (1953);
L.~D.~Landau and I.~Pomeranchuk,
Dokl.\ Akad.\ Nauk Ser.\ Fiz.\  {\bf 92}, 735 (1953);
A.~B.~Migdal,
Phys.\ Rev.\  {\bf 103}, 1811 (1956);
A.~B.~Migdal,
Sov.\ Phys.\ JETP {\bf 5}, 527 (1957).

\bibitem{aharonian}
F.~A.~Aharonian, B.~L.~Kanevsk, and V.~A.~Sahakian, J.\ Phys.\ G {\bf 17}, 199
(1991);
A.~N.~Cillis, H.~Fanchiotti, C.~A.~Garcia Canal and S.~J.~Sciutto,
Phys.\ Rev.\ D {\bf 59}, 113012 (1999)
[arXiv:astro-ph/9809334].


\end{thebibliography}
\end{document}